\documentclass[12pt,reqno]{amsart}
\usepackage{graphicx}
\usepackage{amssymb,amscd,amsbsy}
\usepackage{amsthm}
\usepackage{pdfpages}

\usepackage{amsmath}
\usepackage{amsfonts}   
\usepackage{amssymb} 

\usepackage{graphicx}
\usepackage{amsmath}

\newcommand{\noi}{\noindent}

\def\Rat{\text{\bf Rat}}

\def\ec{\epsilon^{\circ}}

\newcommand{\CP}{{\mathbb C}{\mathbb P}}
\def\PP{\mathcal   P}
\def\[{\left[}
\def\]{\right]}
\def\({\left(}
\def\){\right)}

\newcommand{\eeq}{\end{equation}}
\newcommand{\beq}{\begin{equation}}
\newcommand{\bay}{\begin{eqnarray}}
\newcommand{\ey}{\end{eqnarray}}
\newcommand{\bey}{\begin{eqnarray*}}
\newcommand{\eey}{\end{eqnarray*}}

\newcommand{\R}{\operatorname{res}}

\newtheorem{thm}{\hspace{\parindent}Theorem}[section]

\theoremstyle{remark}

\newtheorem*{rem*}{Remark}

\begin{document}

\newcommand{\vse}{\vspace{.2in}}
\numberwithin{equation}{section}

\title{Analytic  Poisson Brackets  on  rational functions on the Riemann sphere and their applications. }
\author{ K.L. Vaninsky}
\begin{abstract}    We consider a hierarchy of Poisson structures defined on rational  functions on the Riemann sphere. This hierarchy is originated in the theory of the integrable   Camassa-Holm equation associated with the Krein's string spectral problem. Previously the proof of Jacobi identity was obtained  by reducing the bracket to canonical Darboux coordinates. The main result of this note is a direct proof of the Jacobi identity.  It turns out  that the direct proof of the Jacobi identity is  far from trivial. We also give an example of another  hierarchy of Poisson brackets and  construct Darboux coordinates for it.  

   
\end{abstract}
\maketitle

\tableofcontents 
\setcounter{section}{0}
\setcounter{equation}{0}

\section{Introduction}  Consider a space of rational functions  $\Rat_N$ on the Riemann sphere $\CP$ which can be represented as 
$$
w(z)=  \sum_{k=1}^{N} \frac{\rho_k}{z_k-z}=-\frac{q(z)}{p(z)}  ,
$$
Note that the space $\Rat_N$  has complex dimension $2N$. When the parameters $z_k$ and $\rho_k>0$ are real than 
$w(z)$  maps the upper half--plane into itself.  

A simplectic  structure on such space of functions was introduced by Atiyah and Hitchin in \cite{AH} as 
$$
 \sum_{k=1}^{N} \frac{d\, q(z_k)}{q(z_k)} \wedge d\, p(z_k). 
$$
The corresponding  Poisson structure is given by the formula 
\beq\label{ah}
\{w(p), w(q)\} =\frac{\(w(p) - w(q)\)^2}{p-q}. 
\eeq
This form was found in a remarkable paper of Faybusovich and Gekhtman, \cite{FG}.  For the Atiyah-Hitchin bracket \ref{ah} 
it was shown in \cite{V2} that it corresponds  to the main Poisson bracket for  the Camassa-Holm equation written in terms of the Weyl function, 
\cite{W},  of the associated Krein's string spectral problem .

Faybusovich and Gekhtman also found higher brackets  of the infinite hierarchy with 
\ref{ah} being the first bracket.   In our paper KV and Gekhtman, \cite{GV},  we found an algebraic-geometrical representation of all these  brackets.  
In \cite{GV} it was explain that these brackets produce hierarchy of Poisson brackets of the Camassa-Holm equation. 
To introduce a formula for the hierarchy we consider a differential $\alpha_{p \, q}^{f}$ on the Riemann sphere  $\CP$ which depends on the  entire
function $f(z)$ and two points $p$ and $q$ 
$$
\alpha_{p\, q}^f= \frac{ \epsilon_{p q}(z)}{p -q}\times f(z)  w(z) \(w(p)-w(q)\), 
$$
where
$$
\epsilon_{p q}(z)=\frac{1}{2\pi i}\, \[\frac{ 1}{z-p } - \frac{ 1}{z-q}\] dz
$$
is the standard differential Abelian differential of the third kind with  residues $\pm 1$ at the points $p$ and $q$. 
It can be written in the form  
\bey
\alpha_{p\, q}^f&=& \ec_{p q}(z)   
\times f(z)  w(z) \(w(p)-w(q)\)\\
&=&\frac{1}{2\pi i}\, \frac{ d z}{(z-p )\, (z-q)}\times f(z)  w(z) \(w(p)-w(q)\)
\eey
that is used  later. 
The analytic Poisson brackets are defined on $\Rat_N$ by the formula, see \cite{GV}:
\beq\label{apb}
\{w(p),w(q)\}^f= \sum_{k=1}^{N} \int\limits_{\overset{\curvearrowleft}{O}_k} \alpha_{p\, q}^f\; ,
\eeq
where the circles $O_k$ are traversed counter clockwise and surround points $z_k$.  
When $f(z)=1$ we obtain \ref{ah} from \ref{apb}. 
Another closed formula is obtained for 
$f(z)=z$. 

The  bracket \ref{apb}  satisfies the Jacobi identity.  In \cite{GV}   we gave an indirect  proof  which uses the fact that there   exist such  coordinates  that  the bracket \ref{apb} has the  
standard constant form 
\beq\label{conmat}
\newcommand*{\tempi}{\multicolumn{1}{|c}{I}}
\newcommand*{\tempz}{\multicolumn{1}{|c}{0}}
{\mathcal J}=\left[\begin{array}{ccccc}
0      &\tempi \\ \hline
-I      &  \tempz\\
\end{array}
\right].
\eeq
The main result of this note is  a direct proof  of the Jacobi identity  for \ref{apb}.
In fact we give two proofs. One proof presented in Section 2 is based on direct calculations. Another proof given in Section 3 uses the language algebraic geometry. Are there any other hierarchies of Poisson structures? We show in Section 4 that there are only two formulas of the type \ref{ah} 
corresponding to $f(z)=1$ and $f(z)=z$ and that satisfy Jacobi identity. In the last Section 5 we present an example of another hierarchy of Poisson brackets  and construct Darboux coordinates 
for it.

\section{The first proof of Jacobi identity.}

First we give a proof of the Jacobi identity that  use explicit form of  differentials $\ec_{pq}$. 
Everywhere below we omit the superscript $f$ in the formula $\{\;, \;\}= \{\;, \;\}^f$.

\begin{thm}\label{JI}  The bracket defined by \ref{apb} satisfies the Jacobi identity\footnote{$\text{\it c.p.}$  stands for cyclical permutations.} 
$$
\{\{w(p),w(q)\},w(r)\}+ \text{c.p.}=0.
$$
\end{thm}

 {\it Proof.}  From the definition 
$$
\{ w(p), w(q)\} = \frac{1}{2\pi i} \int\limits_{\bigcup O_k}  \frac{  d z \,f(z) w(z) }{(z-p )\, (z-q)}\times   \(w(p)-w(q)\).
$$
Therefore,
\bey
\{\{ w(p), w(q)\},w(r) \}  &=&  \frac{1}{2\pi i} \int\limits_{\bigcup O_k}  \frac{d z\, f(z) }{(z-p )\, (z-q)}\times \{  w(z)\, \(w(p)-w(q)\), w(r) \}\\
&=& \frac{1}{2\pi i} \int\limits_{\bigcup O_k}  \frac{d z\, f(z) w(z) }{(z-p )\, (z-q)}\times \{   \(w(p)-w(q)\), w(r) \} \,  +\\
&+& \frac{1}{2\pi i} \int\limits_{\bigcup O_k}  \frac{ d z\, f(z) }{(z-p )\, (z-q)}\times \{  w(z), w(r) \}\, \(w(p)-w(q)\)\\
 &=& I + II.  \\
\eey

For the first term we have 
\bey
I&=& \frac{1}{2\pi i} \int\limits_{\bigcup O_k}  \frac{  dz f(z)w(z)}{(z-p )\, (z-q)}\times \frac{1}{2\pi i} 
\int\limits_{\bigcup O_{k'}}  \frac{  d \eta f(\eta) w(\eta) }{(\eta-p )\, (\eta-r )} \(w(p)-w(r)\) \\
&-& \frac{1}{2\pi i} \int\limits_{\bigcup O_k}  \frac{ d z f(z) w(z) }{(z-p )\, (z-q)}\times \frac{1}{2\pi i} 
\int\limits_{\bigcup O_{k'}}  \frac{  d \eta f(\eta) w(\eta) }{(\eta-q )\, (\eta-r )} \(w(q)-w(r)\)\\
&=& \frac{1}{(2\pi i)^2 } \int\limits_{\bigcup O_k}   \int\limits_{\bigcup O_{k'}} 
\frac{ dz d\eta\,f(z)f(\eta)  w(z) w(\eta)  (w(p)-w(r)) (z-r) (\eta- q) }  {(  z-p)(z-q)(z-r) (\eta - p)(\eta- r) (\eta-q)}\\
& -& \frac{1}{(2\pi i)^2 } \int\limits_{\bigcup O_k}   \int\limits_{\bigcup O_{k'}} 
\frac{ dz d\eta\, f(z)f(\eta) w(z) w(\eta)  (w(q)-w(r)) (z-r) (\eta- p)}  {(  z-p)(z-q)(z-r) (\eta - p)(\eta- r) (\eta-q)}.
\eey
Denoting $\PP(z)= (  z-p)(z-q)(z-r), $   
\bey
I&=& \frac{1}{(2\pi i)^2 } \int\limits_{\bigcup O_k}   \int\limits_{\bigcup O_{k'}} 
\frac{dz d\eta\, f(z)f(\eta)  w(z) w(\eta)  (w(p)-w(r)) (z-r) (\eta- q) }  {\PP(z) \PP(\eta) }\\
& -& \frac{1}{(2\pi i)^2 } \int\limits_{\bigcup O_k}   \int\limits_{\bigcup O_{k'}} 
\frac{dz d\eta\, f(z)f(\eta)  w(z) w(\eta)  (w(q)-w(r)) (z-r) (\eta- p)}  {\PP(z) \PP(\eta) }
\eey
After simple algebra
\bey
I+ c.p. & =& w(p) (q-r) \frac{1}{(2\pi i)^2 } \int\limits_{\bigcup O_k}   \int\limits_{\bigcup O_{k'}} 
\frac{dz d\eta\,f(z)f(\eta)   w(z) w(\eta) (\eta +p - 2z)  } {\PP(z) \PP(\eta) } \\
   &+&  w(q) (r-p) ... \\
	 &+&  w(r) (p-q) ... .
\eey

Similar for the second term we have 
\bey
II &=& \frac{1}{2\pi i} \int\limits_{\bigcup O_{k'}}  \frac{ d z f(z)}{(z-p )\, (z-q)} \(w(p)-w(q)\) \times \frac{1}{2\pi i} \int\limits_{\bigcup O_k} \frac{d \eta f(\eta) w(\eta)  }{(\eta-z )\, (\eta-r)} \(w(z)-w(r)\) \\
&=& \frac{1}{(2\pi i)^2} \int\limits_{\bigcup O_{k'}} \int\limits_{\bigcup O_k}
\frac{d \eta dz\,f(z)f(\eta)  w(\eta) w(z)\(w(p)-w(q)\) (\eta - p) (\eta-q)(z-r)}{\PP(z) \PP(\eta) (\eta- z)}   \\
&-& \frac{1}{(2\pi i)^2} \int\limits_{\bigcup O_{k'}} \int\limits_{\bigcup O_k} 
\frac{d \eta dz\, f(z)f(\eta) w(\eta) w(r)\(w(p)-w(q)\) (\eta - p) (\eta-q)(z-r) }{\PP(z) \PP(\eta) (\eta- z)} \\
&=& A+ B.   
\eey
From simple algebra 
\bey
B + c.p.&=& w(q)w(p) (p-q) \frac{1}{(2\pi i)^2} \int\limits_{\bigcup O_{k'}} \int\limits_{\bigcup O_k} 
\frac{d \eta dz\,  f(z) f(\eta) w(\eta)  (\eta-r)}{\PP(z) \PP(\eta)} \\
&+& w(p)w(r) (r-p) ...\\
&+& w(r)w(q) (q-r) ...
\eey
Changing the order of integration 
$$
\int\limits_{\bigcup O_{k'}} \int\limits_{\bigcup O_k} 
\frac{d \eta dz\,  f(z) f(\eta)w(\eta)  (\eta-r)}{\PP(z) \PP(\eta)}=
\int\limits_{\bigcup O_{k'}} \frac{d \eta \,f(\eta) w(\eta)  (\eta-r)}{ \PP(\eta)} 
\int\limits_{\bigcup O_k} 
\frac{ dz f(z) }{\PP(z) }.
$$
The differential $dz f(z)/\PP(z) $  is analytic inside the circles $ O_k$ and the integral vanishes due to the Cauchy theorem. 
Therefore, 
$$
B + c.p.=0.
$$
This implies 
\bey
II + c.p.= A +c.p.&=& w(p) (q-r) \frac{1}{(2\pi i)^2 } \int\limits_{\bigcup O_k}   \int\limits_{\bigcup O_{k'}} 
\frac{dz d\eta\, f(z)f(\eta)  w(z) w(\eta) (\eta -p )  } {\PP(z) \PP(\eta) } \\
   &+&  w(q) (r-p) ... \\
	 &+&  w(r) (p-q) ... .
\eey

Finally, 
\bey
I + II + c.p. &=&  \[ w(p) (q-r) +  w(q) (r-p) + w(r) (p-q)\] \times \\
 &\times & \frac{1}{(2\pi i)^2 } \int\limits_{\bigcup O_k}   \int\limits_{\bigcup O_{k'}} 
\frac{dz d\eta\,f(z)f(\eta)   w(z) w(\eta) (2\eta -2z )  } {\PP(z) \PP(\eta) }. 
\eey
The last integral vanishes due to skew symmetry.  \qed

\section{The second proof of Jacobi identity.} 
 
Now we give a proof of the Jacobi identity that  does not use explicit form of  differentials $\ec_{pq}$. 
This proof is a reformulation of direct computations  
above. The proof  uses the language of  algebraic geometry and can be extended to   general spectral curves. 
 
\noindent
 {\it Proof.}  From the definition 
$$
\{ w(p), w(q)\} =  \int\limits_{\bigcup O_k}  \[ \ec_{p q} f\] (z) \times  w(z) \(w(p)-w(q)\).
$$
Therefore,
\bey
\{\{ w(p), w(q)\},w(r) \}  &=&  \int\limits_{\bigcup O_k} \[ \ec_{p q} f\](z) \times \{  w(z)\, \(w(p)-w(q)\), w(r) \}\\
&=& \int\limits_{\bigcup O_k}  \[\ec_{p q} f\](z)\times w(z) \{   w(p)-w(q), w(r) \} \,  +\\
&+&  \int\limits_{\bigcup O_k} \[\ec_{pq} f\](z) \times \(w(p)-w(q)\)\, \{  w(z), w(r) \} \\
 &=& I + II.  \\
\eey

For the first term we have 
\bey
I&=& \int\limits_{\bigcup O_k} \[ \ec_{p q} f w\](z)\times \{ w(p), w(r)\} - 
  \int\limits_{\bigcup O_k}  \[\ec_{p q} f w\](z)\times \{ w(q), w(r)\}\\
&=&  \int\limits_{\bigcup O_k} \[ \ec_{p q} f w\](z)\times 
\int\limits_{\bigcup O_{k'}}   \[ \ec_{p r} f w\](\eta) \times   \(w(p)-w(r)\) \\
& -& \int\limits_{\bigcup O_k} \[ \ec_{p q} f w\](z)\times 
\int\limits_{\bigcup O_{k'}} \[ \ec_{q r} f w\](\eta) \times   \(w(q)-w(r)\)\\
=&+& w(p) \int\limits_{\bigcup O_k} \int\limits_{\bigcup O_{k'}} \[ \ec_{p q} f w\](z)\; \[ \ec_{p r} f w\](\eta)\\
&-& w(r) \int\limits_{\bigcup O_k} \int\limits_{\bigcup O_{k'}} \[ \ec_{p q} f w\](z)\; \[ \ec_{p r} f w\](\eta)\\
&+& w(r) \int\limits_{\bigcup O_k} \int\limits_{\bigcup O_{k'}} \[ \ec_{p q} f w\](z)\; \[ \ec_{q r} f w\](\eta)\\
&-& w(q) \int\limits_{\bigcup O_k} \int\limits_{\bigcup O_{k'}} \[ \ec_{p q} f w\](z)\; \[ \ec_{q r} f w\](\eta). 
\eey
After simple algebra
\bey
I+ c.p. & =& w(p)  \int \int \[ \ec_{p q} f w\](z)\; \[ \ec_{p r} f w\](\eta) -
           \[ \ec_{rp} f w\](z)\; \[ \ec_{p q} f w\](\eta) \\
    &&\qquad\quad\;\;  -\[ \ec_{qr} f w\](z)\; \[ \ec_{qp} f w\](\eta) +\[ \ec_{qr} f w\](z)\; \[ \ec_{rp} f w\](\eta) \\
   &+&  w(q)  ... \\
	 &+&  w(r)  ... .
\eey
Using the first identity
\beq\label{firid}
\frac{\ec_{ab}(z) }{z-c}= \frac{\ec_{a' b'}(z) }{z-c'},  
\eeq
where $(a',b',c')$ is an arbitrary permutation of the points $(a,b,c)$, 
and the second identity
$$
(z-r)(\eta-q)-(z-q)(\eta-r)-(z-p)(\eta-r)+(z-p)(\eta-q)=(\eta+p -2z)(q-r),
$$
we  transform the expression under integral sign to the form
\bey
I+ c.p. & =& w(p) (q-r) \int \int  \frac{\[\ec_{p q}f w\](z)  \[\ec_{p r}f w\](\eta)}{(z-r)(\eta-q)} (\eta+p -2z) \\
              &+&  w(q)  ... \\
	 &+&  w(r)  ... .
\eey

Similar for the second term we have 
\bey
II &=& \(w(p)-w(q)\)\times \,  \int\limits_{\bigcup O_k} \[ \ec_{pq} f\](z)  
\int\limits_{\bigcup O_{k'}}  \[\ec_{z r} f w\](\eta) \times   \(w(z)-w(r)\) \\
&=& \(w(p)-w(q)\)\times \,  \int\limits_{\bigcup O_k}  \[\ec_{pq} f w\] (z)   
\int\limits_{\bigcup O_{k'}}  \[ \ec_{z r} f w\](\eta)    \\
&-& \(w(p)-w(q)\) w(r)\times \,  \int\limits_{\bigcup O_k}  \[ \ec_{pq} f\](z)   
\int\limits_{\bigcup O_{k'}}  \[ \ec_{z r} f w\](\eta)    \\
&=& A-  B.   
\eey

It is easy to see
\bey
A + c.p.&=& w(p) \[ \quad  \int\limits_{\bigcup O_k}   \[\ec_{pq} f w\](z)   
\int\limits_{\bigcup O_{k'}}  \[ \ec_{z r} f w\](\eta) 
 - \int\limits_{\bigcup O_k}   \[\ec_{rp} f\](z)   
\int\limits_{\bigcup O_{k'}}  \[ \ec_{zq} f w \] (\eta)   \]\\
&+& w(q)  ...\\
&+& w(r)  ...  .
\eey
Using the identity
$$
(z-r)(\eta-p)(\eta-q)- (z-q)(\eta-p)(\eta-r)=(\eta - p)(r-q)(z-\eta)
$$
we obtain
\bey
A + c.p.&=& w(p) (r-q)    \int\limits_{\bigcup O_k} \int\limits_{\bigcup O_{k'}}  
\frac{\[\ec_{pq} f w\](z)   \[ \ec_{z r} f w\](\eta)} {(z-r)(\eta-q)} \, (z-\eta) \\
&+& w(q)  ...\\
&+& w(r)  ...  .
\eey
Using the identity
$$
\frac{   \ec_{z r} (\eta)} {\eta-q} \, (z-\eta) =-\ec_{r q}, 
$$
we have 
\bey
A + c.p.&=& w(p) (q-r)    \int\limits_{\bigcup O_k} \int\limits_{\bigcup O_{k'}}  
\frac{\[\ec_{pq} f w\](z)   \[ \ec_{ r q} f w\](\eta)} {(z-r)(\eta-p)} \, (\eta-p) \\
&+& w(q)  ...\\
&+& w(r)  ...  .
\eey

From simple algebra 
\bey
B + c.p.&=& w(q)w(p) \[ \quad  \int\limits_{\bigcup O_k}   \[\ec_{qr} f\](z)   
\int\limits_{\bigcup O_{k'}}  \[ \ec_{z p} f w\](\eta) 
 - \int\limits_{\bigcup O_k}   \[\ec_{r p} f\](z)   
\int\limits_{\bigcup O_{k'}}  \[ \ec_{z q} f w \] (\eta)   \]\\
&+& w(p)w(r)  ...\\
&+& w(r)w(q)  ...
\eey
We are going to transform the expression in the square bracket
using the first identity  \ref{firid} and the second identity 
$$
(z-p)(\eta-q)(\eta-r)- (z-q)(\eta-p)(\eta-r)=(\eta-r)(z-\eta)(p-q). 
$$
Therefore,
\bey
B + c.p.&=& w(q)w(p) 
\times  \[ \quad \int\limits_{\bigcup O_k} \int\limits_{\bigcup O_{k'}} 
\frac{\[ \ec_{qr} f\](z)  \[\ec_{z p} f w\](\eta) (\eta-r)(z-\eta)(p-q)}{(z-p)(\eta-q)(\eta-r)}    
\] \\
&+& w(p)w(r)  ...\\
&+& w(r)w(q)  ...
\eey
Note,
$$
\ec_{z p} (\eta) (z-\eta)=-\frac{d \eta}{\eta -p}=\ec_{p\infty} (\eta).
$$
Changing the order of integration 
\bey
\quad \int\limits_{\bigcup O_k} \int\limits_{\bigcup O_{k'}}&& 
\frac{ \[\ec_{qr} f\](z)  \[\ec_{z p} f w\](\eta) (\eta-r)(z-\eta)(p-q)}{(z-p)(\eta-q)(\eta-r)}  
\\
&=&\int\limits_{\bigcup O_{k'}} \frac{ \[\ec_{p \infty}   f w\](\eta) (\eta-r)(p-q)}{(\eta-q)(\eta-r)}  
\int\limits_{\bigcup O_k} \frac{ \[\ec_{qr} f\](z) }{z-p}  
\eey
The differential  is analytic inside the circles $ O_k$ and the integral vanishes due to the Cauchy theorem. 
Therefore, 
$$
B + c.p.=0.
$$
This implies 
\bey
II + c.p.= A +c.p.&=& w(p) (q-r)    \int\limits_{\bigcup O_k} \int\limits_{\bigcup O_{k'}}  
\frac{\[\ec_{pq} f w\](z)   \[ \ec_{ r q} f w\](\eta)} {(z-r)(\eta-p)} \, (\eta-p)  \\
   &+&  w(q) (r-p) ... \\
	 &+&  w(r) (p-q) ... .
\eey

Finally, 
\bey
I + II + c.p. =  \[ w(p) (q-r) +  w(q) (r-p) + w(r) (p-q)\] \times \\
\times\int\limits_{\bigcup O_k} \int\limits_{\bigcup O_{k'}}  \frac{\[\ec_{pq} f w\](z)   \[ \ec_{ r q} f w\](\eta)} {(z-r)(\eta-p)} \, (2\eta-2z). 
\eey
The last integral vanishes due to skew symmetry.  \qed

\section{The Poisson bracket in terms of the Weyl function.} 

By the Cauchy formula from \ref{apb} we have 
\bey
\{w(p),w(q)\}^f & = & \R_p \, \alpha_{p\, q}^f + \R_q \, \alpha_{p\, q}^f +\R_{\infty} \, \alpha_{p\, q}^f \\
                & = & \frac{f(p) w(p)-f(q)w(q)}{p-q}\(w(p)-w(q)\)+ 
\R_{\infty} \, \alpha_{p\, q}^f.
\eey                
If $f(z)=z^n,\, n=0,1,\hdots;$ then the residue at infinity vanishes identically only for $n=0$ or $1.$ 
The bracket in these cases $n=0$ or $n=1$  has the form 

$$
\{w(p),w(q)\}^1=\frac{w(p)-w(q)}{p-q}\(w(p)-w(q)\),
$$
and 
$$
\{w(p),w(q)\}^z=\frac{p w(p)-q w(q)}{p-q}\(w(p)-w(q)\).
$$
It can be verified directly that these brackets satisfy Jacobi identity. 
Unfortunately these are the only examples of Poisson brackets of such form as the 
following theorem shows.

\begin{thm}
 If $f(z)=z^n,\, n=0,1,\hdots;$ then  the bracket defined as 
$$
\{w(p),w(q)\}^f=\frac{f(p) w(p)-f(q)w(q)}{p-q}\(w(p)-w(q)\),
$$
satisfies the Jacobi identity only for  $n=0$ or $1.$  
\end{thm}

\noi
{\it Proof.} It can be verified in a lengthy  but straightforward computation.

\section{Another Hierarchy of Poisson brackets.}

The new n-th bracket is defined by the formula\footnote{  The formula for $n=0$ was proposed by Philip de Francesco.}  
\beq\label{pb}
\{w(p),w(q)\}^n= p^n w'(p) w(q) - q^n w'(q) w(p). 
\eeq
We omit the  index  $n=0,1,\hdots;$  for the  rest of this section  $\{\;, \; \}=\{\;, \; \}^n$. 

\begin{thm}
The bracket  \ref{pb} satisfies the Jacobi identity
$$
\{\{w(p),w(q)\}, w(r)\}+ \text{c.p.}=0.
$$
\end{thm}

\noi {\it Proof.}
From \ref{pb} we have 
\beq\label{dpb}
\{w'(p),w(q)\}= n p^{n-1}w'(p) w(q)+p^nw''(p) w(q) - q^n w'(q) w'(p). 
\eeq
We compute the first term  in Jacobi identity. Using Leibniz rule 
\bey
\{\{w(p),w(q)\}, w(r)\}&=&\{p^nw'(p) w(q), w(r)\} - \{q^n w'(q) w(p), w(r)\} \\
                                &=& +  p^n w(q) \{w'(p), w(r)\}  +  p^nw'(p)    \{ w(q), w(r)\}    -\\  
																&& - q^n  w(p)\{w'(q), w(r)\} - q^n w'(q) \{w(p), w(r)\}.    \\
\eey
Applying \ref{dpb} 
\bey 
       \hdots &=& +p^n w(q) n p^{n-1}  w'(p) w(r)+p^n w(q)  p^n w''(p) w(r) -p^n w(q) r^n w'(r) w'(p)\\
				&& +p^n w'(p) q^n w'(q) w(r) -p^n w'(p) r^n w'(r) w(q)\\
				&& -q^n w(p) n q^{n-1} w'(q) w(r)-q^n w(p)  q^n w''(q) w (r) +q^n w (p) r^n w '(r) w '(q)\\
				&& -q^n w '(q) p^n w '(p) w (r) +q^n w '(q) r^n w (p)  w '(r). 
\eey
After a few cancellations the first term  becomes 
\bey 
\{\{w (p),w (q)\}, w (r)\}&=&  + p^{2n}w (q)  w ''(p) w (r)-q^{2n}w (p)  w ''(q) w (r)\\
                                && -2 p^n r^n w '(p)  w '(r) w (q) +2 q^n r^n w '(q) w '(r) w (p)\\
																&& + n p^{2n-1}w (q)  w '(p) w (r)-  nq^{2n-1}w (p)  w '(q) w (r). 
\eey 
Circular permutations of variables produce 

\bey 
\{\{w (r),w (p)\}, w (q)\}&=&  + r^{2n}w (p)  w ''(r) w (q)-p^{2n}w (r)  w ''(p) w (q)\\
                                && -2 r^n q^n w '(r)  w '(q) w (p) +2 p^n q^n w '(p) w '(q) w (r)\\
																&& + n r^{2n-1}w (p)  w '(r) w (q)-  n p^{2n-1}w (r)  w '(p) w (q). 
\eey 
and 
\bey 
\{\{w (q),w (r)\}, w (p)\}&=&  + q^{2n}w (r)  w ''(q) w (p)-r^{2n}w (q)  w ''(r) w (p)\\
                                && -2 q^n p^n w '(q)  w '(p) w (r) +2 r^n p^n w '(r) w '(p) w (q)\\
																&& + n q^{2n-1}w (r)  w '(q) w (p)-  n r^{2n-1}w (q)  w '(r) w (p). 
\eey 
The sum of three brackets is zero. \qed

Now we compute using methods of \cite{V1} Darboux coordinates for the bracket $\{\;, \; \}^n$.  

\begin{thm} 
For the bracket \ref{pb} the following  identities hold
\bay
\{\rho_k,\rho_p\}&=&-\rho_k \rho_p n(z_k^{n-1}-z_p^{n-1}), \label{fb}\\
\{\rho_p, z_k \} &=& \rho_p z_k^n , \label{sb} \\
\{z_k, z_p\} &=&0. \label{tb} 
\ey
\end{thm}

\noi {\it Proof.}
Note that 
$$
\rho_k= \frac {1}{2 \pi i } \int\limits_{O_k} w (\zeta) d\zeta, \qquad\qquad\qquad \rho_k  z_k= \frac {1}{2 \pi i } \int\limits_{O_k} \zeta 
w (\zeta) d\zeta. 
$$
Therefore,  integrating by parts 
$$
\frac {1}{2 \pi i } \int\limits_{O_k} d\zeta\, \zeta^n w '(\zeta)= - \frac {1}{2 \pi i } \int\limits_{O_k} d\zeta\, n\,\zeta^{n-1} w (\zeta)
= -n z_k^{n-1}\rho_k.
$$

To prove \ref{fb} we have 
\bey
\{\rho_k,\rho_p\}&=&\{  \frac {1}{2 \pi i } \int\limits_{O_k} w (\zeta) d\zeta    , \frac {1}{2 \pi i } \int\limits_{O_p} w (\eta) d\eta  \} \\
&=& \frac {1}{(2 \pi i)^2 } \int\limits_{O_k}\int\limits_{O_p}d\zeta d\eta \[ \zeta^n w '(\zeta)  w (\eta) - \eta^n w '(\eta) w (\zeta)\] \\ 
&=& \frac {1}{2 \pi i } \int\limits_{O_k} d\zeta\, \zeta^n w '(\zeta) \frac {1}{2 \pi i } \int\limits_{O_p}d\eta w (\eta) - 
    \frac {1}{2 \pi i } \int\limits_{O_p} d\eta\, \eta^n w '(\eta) \frac {1}{2 \pi i } \int\limits_{O_k}d\zeta w (\zeta)\\
		&=& -\rho_k n z_k^{n-1}  \rho_p + \rho_k n z_p^{n-1} \rho_p. 
\eey

To prove \ref{sb} we have 
\bey
\{z_k \rho_k,\rho_p\}&=&\{  \frac {1}{2 \pi i } \int\limits_{O_k} \zeta w (\zeta) d\zeta    , \frac {1}{2 \pi i } \int\limits_{O_p} w (\eta) d\eta  \} \\
&=& \frac {1}{(2 \pi i)^2 } \int\limits_{O_k}\int\limits_{O_p}d\zeta d\eta \[ \zeta^{n+1} w '(\zeta)  w (\eta) - \eta^n w '(\eta) \zeta w (\zeta)\] \\ 
&=& \frac {1}{2 \pi i } \int\limits_{O_k} d\zeta\, \zeta^{n+1} w '(\zeta) \frac {1}{2 \pi i } \int\limits_{O_p}d\eta w (\eta) - 
    \frac {1}{2 \pi i } \int\limits_{O_p} d\eta\, \eta^n w '(\eta) \frac {1}{2 \pi i } \int\limits_{O_k}d\zeta \zeta w (\zeta)\\
		&=& -\rho_k (n+1) z_k^{n}  \rho_p +  z_k \rho_k n z_p^{n-1} \rho_p. 
\eey
By Leibnitz rule
\bey
-\rho_k (n+1) z_k^{n}  \rho_p +  z_k \rho_k n z_p^{n-1} \rho_p&=&\{z_k \rho_k,\rho_p\}= z_k \{ \rho_k,\rho_p\} +\rho_k \{z_k ,\rho_p\}\\
                  &=&- n z_k \rho_k   \rho_p (z_k^{n-1} - z_p^{n-1} )  +  \rho_k \{z_k ,\rho_p\}. 
\eey
Therefore,
$$
\rho_k \{z_k ,\rho_p\}= -\rho_k (n+1) z_k^{n}  \rho_p  + n  \rho_k   \rho_p z_k^{n}= - \rho_k   \rho_p z_k^{n}. 
$$

To prove \ref{tb} we have 
\bey
\{z_k \rho_k,z_p \rho_p\}&=&\{  \frac {1}{2 \pi i } \int\limits_{O_k} \zeta w (\zeta) d\zeta    , \frac {1}{2 \pi i } \int\limits_{O_p} \eta w (\eta) d\eta  \} \\
&=& \frac {1}{(2 \pi i)^2 } \int\limits_{O_k}\int\limits_{O_p}d\zeta d\eta \[ \zeta^{n+1} w '(\zeta) \eta w (\eta) - \eta^{n+1}  w '(\eta) \zeta w (\zeta)\] \\ 
&=& \frac {1}{2 \pi i } \int\limits_{O_k} d\zeta\, \zeta^{n+1} w '(\zeta) \frac {1}{2 \pi i } \int\limits_{O_p}d\eta \eta w (\eta) - 
    \frac {1}{2 \pi i } \int\limits_{O_p} d\eta\, \eta^{n+1}  w '(\eta) \frac {1}{2 \pi i } \int\limits_{O_k}d\zeta \zeta w (\zeta)\\
		&=& -\rho_k (n+1) z_k^{n} z_p \rho_p +   \rho_p (n+1) z_p^{n}  z_k \rho_k . 
\eey
Again, by Leibnitz rule
\bey
-\rho_k (n+1) z_k^{n} z_p \rho_p &+&   \rho_p (n+1) z_p^{n}  z_k \rho_k =\{z_k \rho_k,z_p \rho_p\} \\
&=& \rho_k \rho_p \{z_k ,z_p \}+ \rho_k z_p \{z_k , \rho_p\}
+z_k \rho_p \{\rho_k,z_p \}+ z_k z_p \{ \rho_k,\rho_p\}\\
&=& \rho_k \rho_p \{z_k ,z_p \}- \rho_k z_p z_k^n  \rho_p
+z_k \rho_p\rho_kz_p^n   - z_k z_p \rho_k \rho_p n(z_k^{n-1}-z_p^{n-1})
\eey
This implies \ref{tb}. \qed

Let $n=0$. Then we have
$$
\{\rho_k,\rho_p\}=\{z_k, z_p\} =0,\qquad  \{\rho_p, z_k \} = \rho_p.  
$$
To reduce the bracket to constant form define
$$
I_k=z_k,\qquad\qquad \theta_k=\log \rho_k. 
$$
Then the Poisson tensor in $I-\theta$ coordinates has the form 

$$
\newcommand*{\tempo}{\multicolumn{1}{|c}{\bf 1}}
\newcommand*{\tempz}{\multicolumn{1}{|c}{\bf 0}}
{\mathcal J}=\left[\begin{array}{ccccc}
\bf 0      &\tempo \\ \hline
-\bf 1     &  \tempz\\
\end{array}
\right],
$$
where $\bf 0$ the $N\times N$ zero matrix and $\bf 1$ is the $N\times N$ matrix with all entries equal to $1$. 
The Poisson bracket is highly degenerate and has rank  $2$, see \cite{WE}. Let 
$$
\mathcal I= \frac{I_1+I_2+\hdots + I_N}{N},
$$
and 
$$
 \Theta= \frac{\theta_1+\theta_2+\hdots + \theta_N}{N}. 
$$
Then,
$$
\{ \mathcal I, \Theta \}=1. 
$$
The matrix $\bf 1$  has rank  $N-1$ and it is easy to construct $2N-2$ linear functions 
$$
C_1, C_2, \hdots , C_{2N-2};
$$
which are Casimirs of the bracket.

Let $n=1$. Then we have
$$
\{\rho_k,\rho_p\}=\{z_k, z_p\} =0,\qquad  \{\rho_p, z_k \} = \rho_p z_k.  
$$
Again, lets us introduce new variables 
$$
I_k=\log z_k,\qquad\qquad \theta_k=\log \rho_k. 
$$
The Poisson tensor for these variables is the matrix $\mathcal J.$

For $n=2,3,...;$ we define 
$$
I_k=z_k^{-n+1},\qquad\qquad \theta_k=\log \rho_k. 
$$
The Poisson tensor is the matrix $\mathcal J$  and the analysis above can be applied.

\vskip .2in
\noindent

Kirill Vaninsky
\newline
Department of Mathematics
\newline
Michigan State University
\newline
East Lansing, MI 48824
\newline
USA
\noindent
\newline
vaninsky@math.msu.edu

\newpage

\end{document}